\documentclass[twocolumn,aps,prb,floatfix,showpacs]{revtex4-1}
\usepackage{graphicx}
\usepackage{graphics}
\usepackage{amssymb,amsmath} 
\usepackage{dcolumn}
\usepackage{bm}
\usepackage{subfigure}
\usepackage[colorlinks,bookmarks=false,citecolor=red,linkcolor=blue,urlcolor=blue]{hyperref}

\def \S{{\bf S}}

\begin{document}
\title{DMRG studies on linear-exchange quantum spin models in one dimension}
\author{Bimla Danu}
\email{danubimla1@gmail.com}
\author{Brijesh Kumar}
\email{bkumar@mail.jnu.ac.in} 
\affiliation{School of Physical Sciences, Jawaharlal Nehru University, New Delhi 110067, India}
\author{Ramesh V. Pai}
\affiliation{Department of Physics, Goa University, Goa 403206, India}
\date{\today}
\begin{abstract}
We study a class of spin-$1/2$ quantum antiferromagnetic chains using DMRG technique. The exchange interaction in these models decreases linearly as a function of the separation between the spins, $J_{ij} = R-|i-j|$ for $|i-j| \le R$. For the separations beyond $R$, the interaction is zero. The range parameter $R$ takes positive integer values. The models corresponding to all the odd values of $R$ are known to have the same exact doubly degenerate dimer ground state as for the Majumdar-Ghosh (MG) model. In fact, $R=3$ is the MG model. For even $R$, the exact ground state is not known in general, except for $R=2$ (the Bethe ansatz solvable Heisenberg chain) and in the asymptotic limit of $R$ where the two MG dimer states again emerge as the exact ground state. In the present work, we numerically investigate the even-$R$ models whose ground state is not known analytically. In particular, for $R=4$, 6 and 8, we have computed a number of ground state properties. We find that, unlike $R=2$, the higher even-$R$ models are spin-gapped, and show strong dimer-dimer correlations of the MG type. Moreover, the spin-spin correlations decay very rapidly, albeit showing weak periodic revivals.
\end{abstract}
\maketitle
 \section{Introduction:} The quantum spin chains have been a subject of intense research effort for many decades.\cite{1dQM,Haldane1983,Osterloh2002} While the quantum fluctuations manifest strongly in one dimension, the frustration also brings out interesting possibilities such as the spontaneous dimerization as found for the first time in the ground state of the Majumdar-Ghosh (MG) model.\cite{MG} Apart from the theoretical interests initiated by the Bethe's solution of the nearest-neighbor Heisenberg chain,\cite{bethe1931} or by the exact dimer ground state of the MG model, there exist a variety of (quasi) one dimensional materials which provide strong reasons for studying one-dimensional spin problems.\cite{CuGeO31993,LiCuSbO42012,Sr2CuPo42006} Although quite a few of the one-dimensional quantum spin models are known to admit full or partial exact analytical solution, in general, one still depends upon numerical approaches to investigate these problems. The exact numerical diagonalization has been the most standard method, but it is severely limited by the size. \cite{Lin1990,Sandvik}  In the wake of this, the density matrix renormalization group (DMRG) method has emerged as a valuable numerical tool of computation for accessing large system sizes.\cite{WhiteNoack1992,white1993,Schollwock2005} Although approximate, the DMRG is an excellent technique to study the ground state and the lowest few excited states of one dimensional model Hamiltonians.\cite{WhiteHuse1993,HRK1996,Bursill1996,Hallberg1995}  
 
The present work is aimed at studying, using DMRG, a class of one-dimensional quantum spin-1/2 antiferromagnetic models in which the exchange interaction varies linearly with the spin-spin distance upto a distance of $R-1$, beyond which it is zero. The Hamiltonian for this class of `linear-exchange' models is written as follows.
\begin{equation}
H^{ }_R = \sum_i\sum_{n=1}^{R-1} (R-n)\S_i\cdot\S_{i+n}
\label{eq:HR}
\end{equation}
Here, $i$ is the lattice-site label, and $R$ takes positive integer values $\ge 2$, and generates different models. The construction of $H^{ }_R$ in Refs.~\onlinecite{bkumar2002,bkumar-thesis} was inspired by the structure underlying the MG model. Interestingly, the first of these models, $H_2$ for $R=2$, is the famous nearest-neighbor Heisenberg problem solvable by Bethe ansatz.  The next model $H_3$, for $R=3$, is in fact the MG model whose exact doubly degenerate ground state is given by the following singlet-dimer configurations.
\begin{subequations}
\begin{eqnarray}
&& |MG1\rangle= [1,2\rangle\otimes[3,4\rangle\otimes\cdots\otimes [L-1,L\rangle \label{eq:MG1}\\
&& |MG2\rangle= [2,3\rangle\otimes[4,5\rangle\otimes\cdots\otimes [L,1\rangle \label{eq:MG2}
\end{eqnarray}
\end{subequations}
Here, $[i,j\rangle=\left(|\uparrow_i \downarrow_j\rangle-|\downarrow_i \uparrow_j\rangle\right)/\sqrt{2}$ denotes the singlet state formed by the spins at sites $i$ and $j$. In writing the two MG states above, it is assumed that we have an even number of total spins, $L$, and a lattice with periodic boundary condition. 

Interestingly, the doubly degenerate dimer ground state of the MG model, $H^{ }_3$, is also known to be the exact ground state of $H^{ }_R$ for any odd value of $R$. Thus, all the odd-$R$ linear-exchange models are exactly solvable for the ground state.\cite{bkumar2002,takano1} It is also known that the elementary excitations, which can be viewed as dispersing spin-1/2 domain walls (spinons),\cite{SS-spinon} are gapped.\cite{bkumar-thesis, takano2} The situation for the even-$R$ models is different, however. The $H^{ }_2$ happens to be the famous integrable model solvable by Bethe ansatz. It is known to have algebraic spin-spin correlations in the ground state and gapless excitations. Apart from $H_2$, the only other rigorously understood case (at least for the ground state) is the  asymptotic limit of $R$, defined as $R\sim \mathcal{O}(L/2)$ and $L\rightarrow\infty$ ($L$ being the total number of spins on the chain). In this asymptotic limit, the doubly degenerate MG ground state has been shown to become the exact ground state of the even-$R$ models.\cite{bkumar-thesis} Except for $H^{ }_2$ and the asymptotic limit, currently, there is no rigorous understanding of the even-$R$ linear-exchange models. 

The main objective of the present work is to investigate the even-$R$ linear exchange models for finite $R\ge4$. We have used DMRG technique to numerically study the ground state properties of these models, in particular the $H^{ }_4$, $H^{ }_6$ and $H^{ }_8$. We find that $H_{R\ge4}$ behave qualitatively differently from $H_2$, quite unlike the odd-$R$ models, all of which behave like $H_3$. For example, the elementary excitations in $H_2$ are known to be gapless, but we find the even-$R\neq2$ models to be spin-gapped. Correspondingly, the spin-spin correlation in their  ground states decays very rapidly. Moreover, they exhibit the MG type dimer order in the ground state which strongly tends, with increasing $R$, towards the asymptotic behavior. Thus, we find the even-$R\neq 2$ models standing closer to the odd-$R$ models, while the $H_2$ stands alone. This is contrary to a naive expectation that all the even-$R$ models might behave qualitatively similarly, as the odd-$R$ models do. Although the spin-spin correlations decay rapidly for the even-$R\neq2$ models, they exhibit weak periodic revivals. We analyze this in some detail, and find ($\frac{R}{2}-1$) prominent wavenumbers, corresponding to the correlation-revival, pretty close to the integer multiples of $2\pi/R$ (in addition to $\pi$, which is the dominant antiferromagnetic wavenumber anyway), as expected from an Ising type analysis.\cite{bkumar-thesis}

This paper is organized as follows.  In Sec.~\ref{dmrg}, we briefly discuss the DMRG algorithm and the computation of various physical quantities within DMRG. In Sec.~\ref{discussion}, we present the results and discussion of the ground state properties of $H_4$, $H_6$ and $H_8$. Finally, we conclude with a summary.   
\section{Quick overview of DMRG}\label {dmrg}
The DMRG is a very useful method which allows us to calculate a lot of useful quantities, such as the ground state energy, the energy gap and the spin-spin correlations, for large system sizes. Operationally speaking, it is an efficient numerical scheme for truncating the Hilbert space of low dimensional (especially one dimensional) quantum systems. The basic idea is to optimize the basis states by keeping a bunch of highly probable eigenstates of the density matrix at successive stages of renormalization.\cite{Schollwock2005} 

The implementation of DMRG uses a superblock structure consisting of a system and an environment. For the system states denoted as $|i\rangle$, and the environment states denoted as $|j\rangle$, the superblock ground state, $|\psi \rangle$, can be written as: $|\psi\rangle=\sum_{i,j} \psi_{ij}|i\rangle|j\rangle$, where $i$ and $j$ are summed over the complete states of the system and environment, respectively. The ground state density matrix of the superblock is defined as $\rho^{ }_{0}=|\psi\rangle\langle\psi|$. The reduced density matrix, $\rho$, of the system block is obtained by tracing $\rho^{ }_{0}$ over the environment. It can be written as: $\rho_{ii^\prime} = \sum_j \psi_{ij}\psi^*_{i^\prime j}$. Here, $\rho_{ii^\prime}$ is the matrix element $\langle i|\rho|i^\prime\rangle$ of the reduced density matrix of the system. The $\rho$ play a crucial role in DMRG, because a selection of its large eigenvalue states forms the truncated basis for constructing a renormalized superblock.

There are two different DMRG algorithms in use, the infinite and the finite system algorithms, differing by the choice of the environment block in relation to the system block.\cite{white1993}
In the infinite system algorithm, the environment is the mirror image of the system. The starting superblock consists of four (or more) bare sites, of which the two on the left-hand-side may be called the system and the remaining two sites define the environment. Both the blocks grow by one site at a time, increasing the superblock size by two sites at each iteration. Since, in principle, there is no limit on this growth, it is called the infinite system DMRG. The iterations are repeated until the desired number of sites is reached. 

The most important step in each iteration is to truncate the system basis by keeping only a certain number of high probability eigenstates of $\rho$. The system block in the truncated basis acts as an updated left-most site, and its mirror image as the right-most site. Adding two new bare sites between them forms a renormalized superblock and completes one iteration of DMRG. It is important that the truncated basis dimension is not too small to give poor accuracy. If we choose $m$ highest eigenstates, then a measure of the truncation error is the weight of the discarded states, $p_m=1-\sum^{m}_{i=1}\omega_{i}$, where $\omega_{i}$'s are the eigenvalues of $\rho$. Naturally, keeping more states is always better for accuracy. But unfortunately $m$ can not be very large either, because the superblock's Hilbert space grows exponentially which makes the computation very hard.

In the finite system algorithm, first one uses the infinite system algorithm to build up a superblock of the desired size $L$. After that, at each RG step, the system is made to grow at the expense of the environment (or vice versa), keeping the total length $L$ fixed. Now, the system and the environment blocks are asymmetrical in general. One needs to do a few sweeps (typically three or four) of growing and diminishing system size to obtain more accurate results (specially when the system is gapless).

For the DMRG calculations on the $H_R$ models, where the interactions go beyond the nearest neighbors, we construct the superblocks in such a manner that the two renormalized blocks on the left and right do not directly interact with each other. This is done to avoid the density matrix errors. Hence, we have taken four bare sites between the left and right blocks for $R=3, 4, 5$, six bare sites in the middle for $R=6, 7$, and eight bare sites for $R=8, 9$. We have used open boundary condition throughout our calculations.  We have also implemented the total $S_z$ symmetry, and solved $H_4$ with $m=200$, $H_6$ with $m=150$, and $H_8$ with $m=95$. We find that, in the total $S_z=0$ sector for $R=4$, the truncation error is of the order of $10^{-13}$. For $H_6$, it is $\sim 10^{-10}$, and for $H_8$ it is around $10^{-8}$. In total $S_z=1$ sector, for $H_4$ it is $\sim 10^{-10}$; for $R=6$, it is $\sim 10^{-8}$; and, for $R=8$ it is of the order of $10^{-6}$. For the odd-$R$ models, the typical truncation errors are of the order of $10^{-15}$ in the total $S_z=0$ sector. In total $S_z=1$ sector, for $R=3$, it is $\sim 10^{-13}$; for $R=5$, it is $\sim 10^{-10}$; for $R=7$, it is $\sim 10^{-9}$; and, for $R=9$ it is of the order of $10^{-7}$.


 \section{Results and discussion}\label {discussion}
 We now present the results of DMRG calculations for the linear-exchange spin models. In the calculations for a given $H_R$, the nearest neighbor interaction, $R-1$, is set as the unit of energy. To characterize the even-$R$ models, we have computed a number of useful quantities. Below we present them one by one, and discuss their implications for the nature of these models. We have performed calculations on the models upto $R=8$. Going beyond it becomes computationally difficult. In any case, this is sufficient to get a very clear picture of the even-$R\neq 2$ linear-exchange models.
 \subsection{Ground state energy}
\begin{figure}[h]
\centering
 \includegraphics[width=8.25cm]{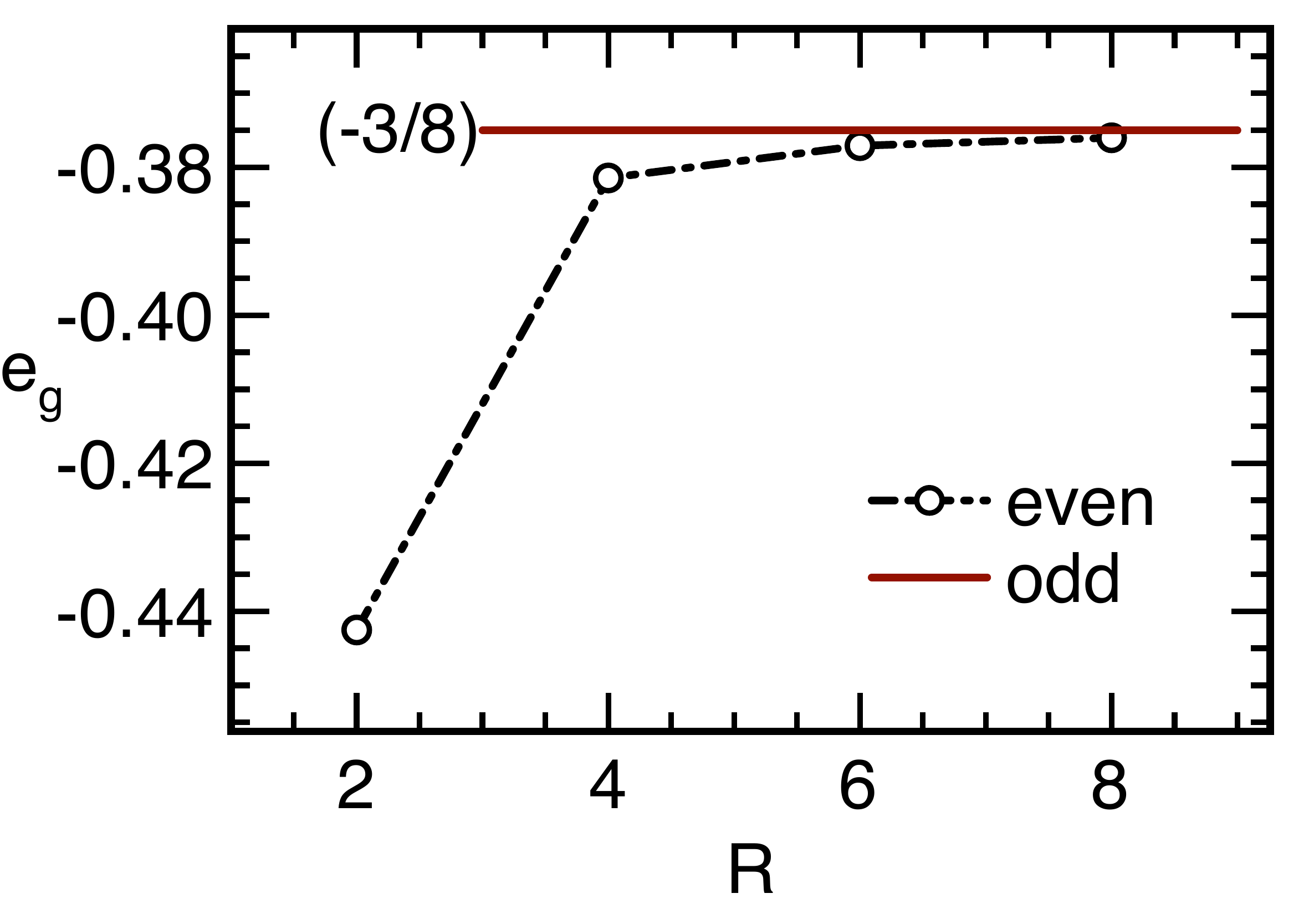}
\caption { Ground state energy per site vs. $R$ for $L=300$.}
 \label{fig:Eg}
 \end{figure}

 We benchmark our DMRG codes by computing the exactly known ground state energies (per spin), $e_g$, of the $H_2$ and the odd-$R$ models. Of course, we get the correct values of  $\frac{1}{4}-\log{2}=-0.443147$ for $H_2$, and -3/8 for the odd-$R$ models. The $e_g$ for $R=2, 4, 6$ and 8 are presented in Fig.~\ref{fig:Eg}, together with -3/8 for the odd-$R$. It is clear that the $e_g$ for $R \neq 2$ rapidly tends to -3/8. In fact, for $R=8$, it is already very close. The data is in agreement with the exact diagonalization results on small clusters, and clearly approaches the exact asymptotic limit.\cite{bkumar-thesis} 
 \subsection{Spin-gap}
\begin{figure}
\centering
\includegraphics[width=8cm]{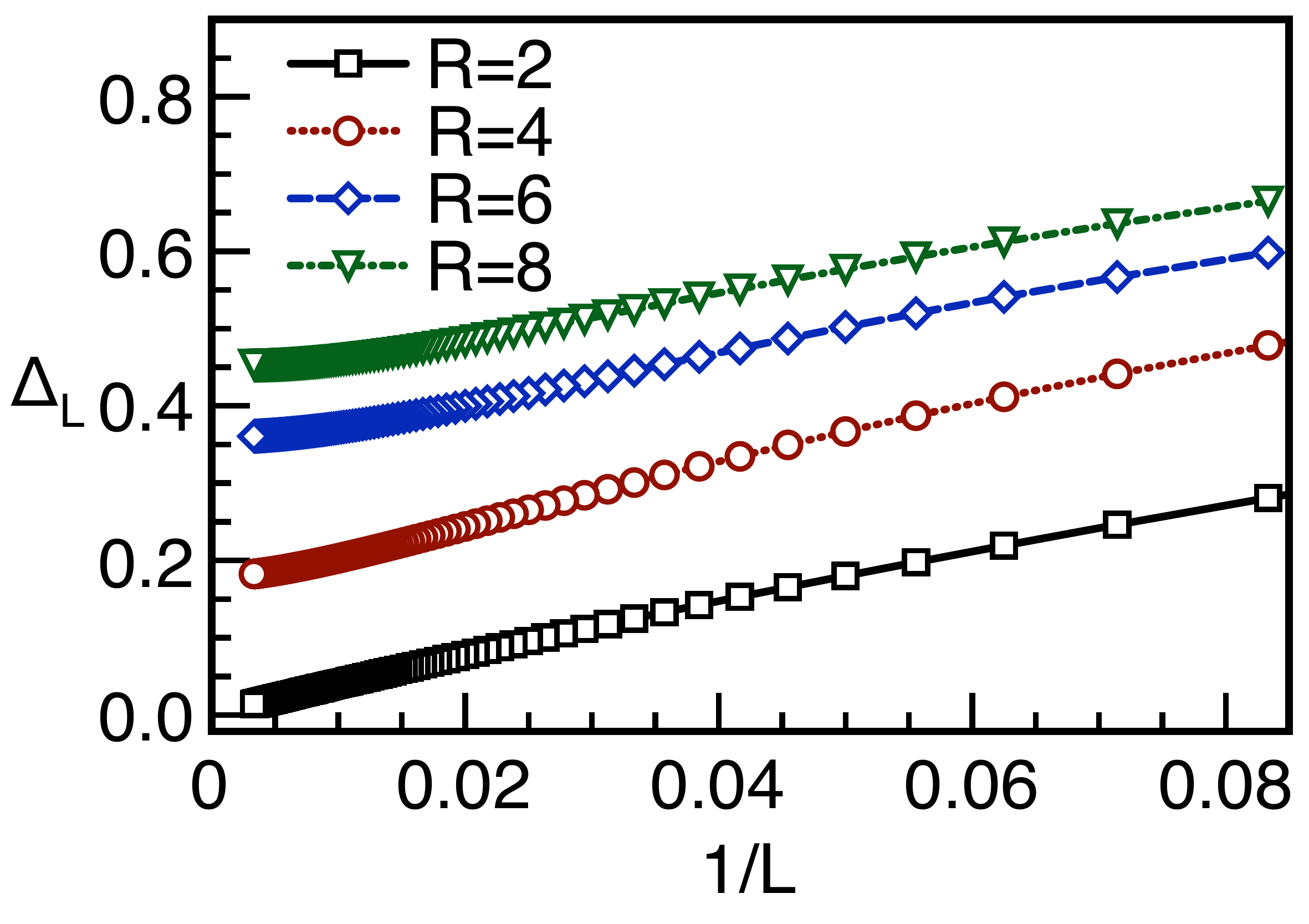}
\caption {Spin-gap vs. $1/L$  up to  $L=300$.}
\label{fig:gap}
\end{figure}

The odd-$R$ linear-exchange models are known to be spin-gapped. Also known is the gapless behavior of $H_2$. The challenge is to find out if the even-$R\neq 2$ models are gapped or gapless. It is not obvious a-priori, although the behavior of $e_g$ in Fig.~\ref{fig:Eg} suggests that, for large enough $R$, the even-$R$ models may show up gapped behavior. To our surprise, we find a robust non-zero energy gap for an $R$ as small as 4, as shown in Fig.~\ref{fig:gap}. The gap increases with increasing $R$. The thermodynamic values of the spin-gaps are estimated to be $\Delta_\infty^{R=4} \approx 0.17$, $\Delta_\infty^{R=6} \approx 0.35$ and $\Delta_\infty^{R=8} \approx 0.45$. The energy gap to spin excitations is estimated by computing $\Delta_L=E_g^{S_z=1}-E_g^{S_z=0}$ for different values of $L$, and extrapolating to $1/L=0$. Here, $E_g^{S_z=0}$ and $E_g^{S_z=1}$ are the lowest energy eigenvalues in the total $S_z=0$ and 1 sectors, respectively.

According to the Lieb-Schultz-Mattis (LSM) theorem for spin-1/2 chains, the existence of  an energy gap in the thermodynamic limit ($L\rightarrow\infty$) implies a (doubly) degenerate ground state which is typically facilitated by an enlarged (dimerized) unit cell. The odd-$R$ linear-exchange models clearly satisfy the LSM theorem. So does $H_2$. The LSM actually proved for $H_2$ a unique ground state and gapless excitations.\cite{LSM} In view of the LSM theorem, the existence of spin-gap in the even-$R\neq 2$ models strongly suggests dimerization in the thermodynamic limit, even for finite $R$. It doesn't mean that the ground wavefunctions for these models will be same as the exact MG states. Nonetheless, it is expected that they show similar singlet-dimer order as in the MG state, which in the asymptotic limit is known to become exact.\cite{bkumar-thesis}
 \subsection{Dimer order}
 \begin{figure}
\centering
\includegraphics[width=8cm]{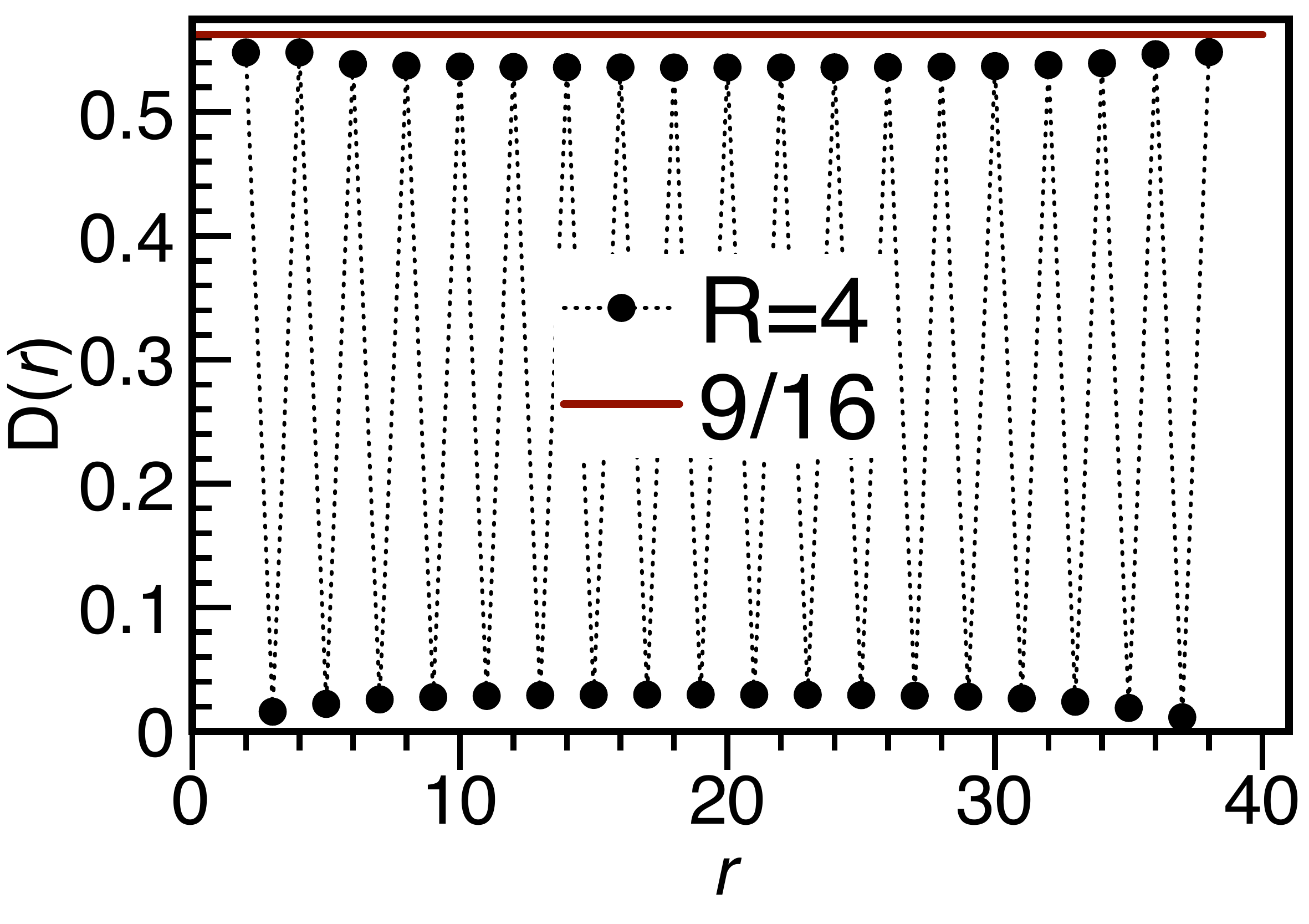}\\
\includegraphics[width=8cm]{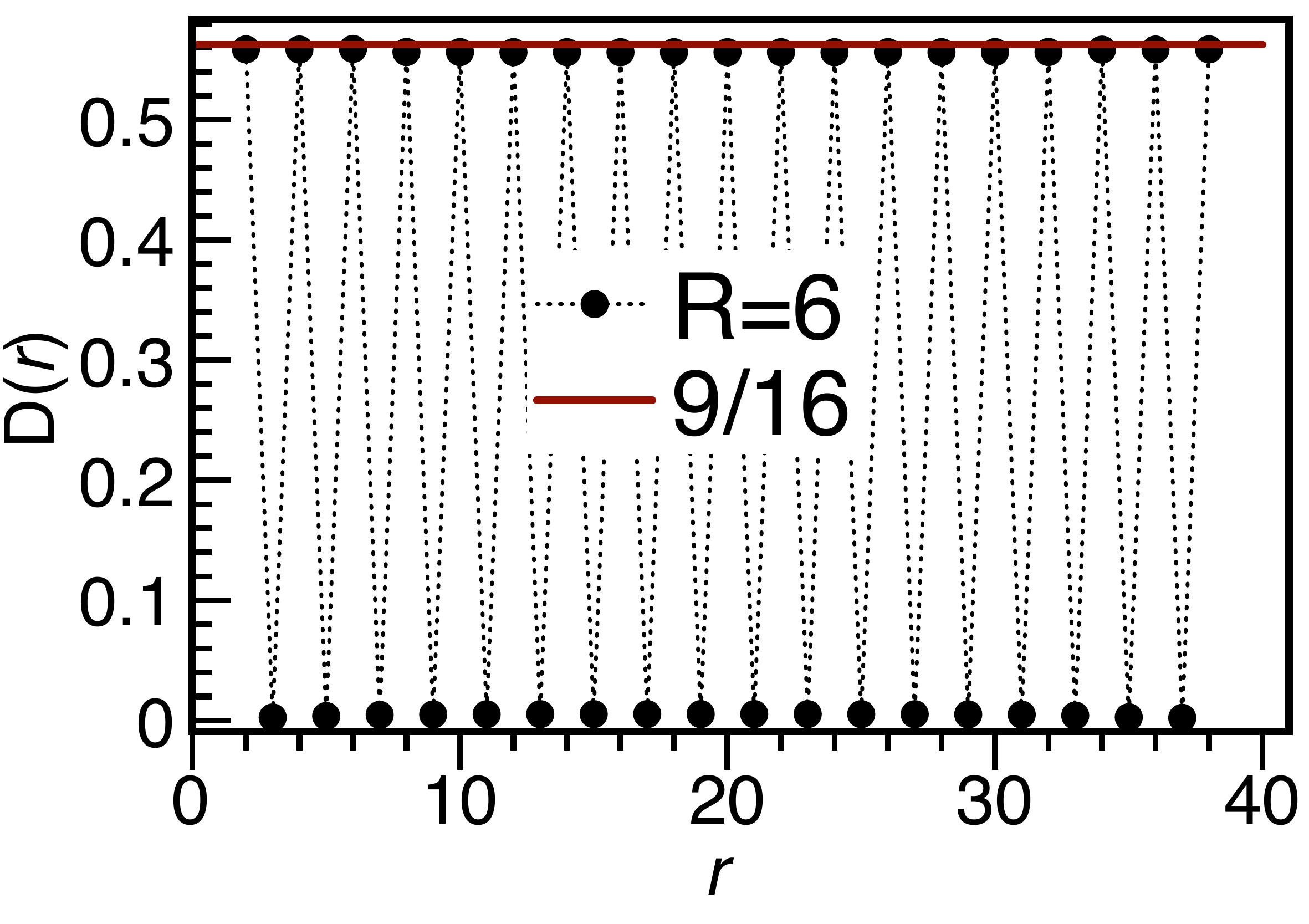}\\
\includegraphics[width=8cm]{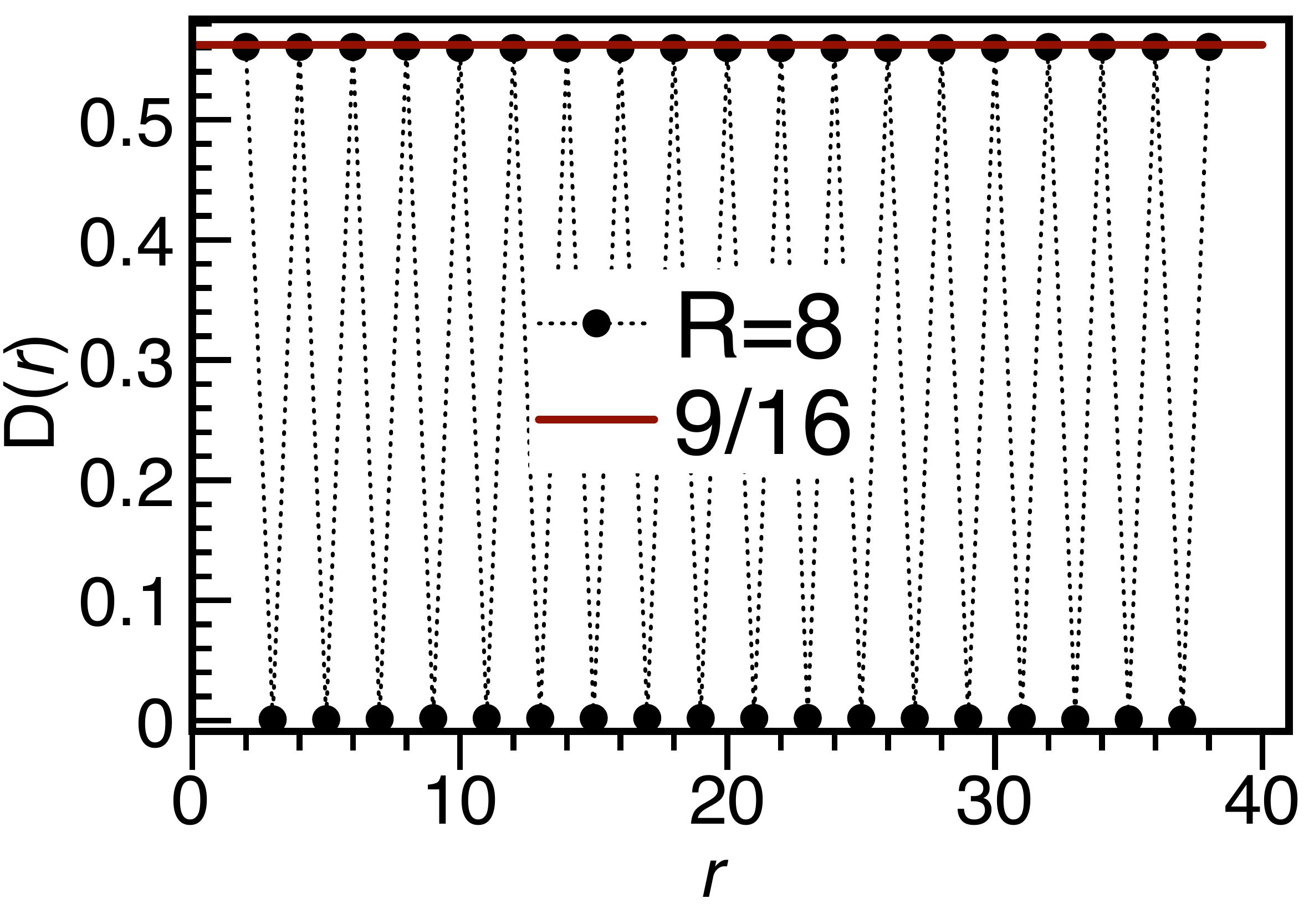}
\caption { Dimer-dimer correlation for $L=40$. }
\label{fig:R468ddcr}
\end{figure}
 Guided by the LSM theorem, and the asymptotic limit, we investigate dimerization in the ground states of $H_4$, $H_6$ and $H_8$. The results of DMRG calculations for the dimer-dimer correlation, $D(r)=\langle \S_{1}\cdot\S_{2}~\S_{1+r}\cdot\S_{2+r}\rangle$, are shown in Fig.~\ref{fig:R468ddcr}. The data computed for $L=40$, 60 and 80 essentially looks the same. The $D(r)$ measures the correlation between the total-spin states on different nearest-neighbor bonds. If two such bonds are perfect singlets, then $D$ is equal to $(-3/4)^2$=9/16 (=0.5625). For the MG states, $D(r)$ switches between 0 and 9/16 every time $r$ increases by one, because every alternate pair of nearest-neighbor spins forms a singlet. Notably,  for $R=4$, 6 and 8 too, the same qualitative behavior is found. Quantitatively, for $H_4$, the minimum of $D(r)$ is quite close to zero and the maximum is only slightly less than 9/16. For $R=6$, the numbers get pretty close to the ideal values. For $H_8$, the $D(r)$ is almost exactly like in the MG ground state. This data clearly demonstrates the nearest-neighbor singlet formation in the ground states of the even-$R\neq 2$ models, and puts them qualitatively alongside the odd-$R$ models. 
 
 It should be noted that on an open boundary chain, for the odd-$R$ models, of the two degenerate MG states in Eqs.~(\ref{eq:MG1}) and~(\ref{eq:MG2}), only $|MG1\rangle$ is selected as the ground state. So, while the nature of the ground state remains unchanged, the degeneracy is `artifically' changed to one. The same is happening for the even-$R\neq 2$ models. We are getting the MG dimer order in the ground state, but because of the open boundary condition of the DMRG computations, we do not get two-fold degeneracy.  
 
 \subsection{Spin-spin correlation}
 \begin{figure}
\centering
\includegraphics[width=8cm]{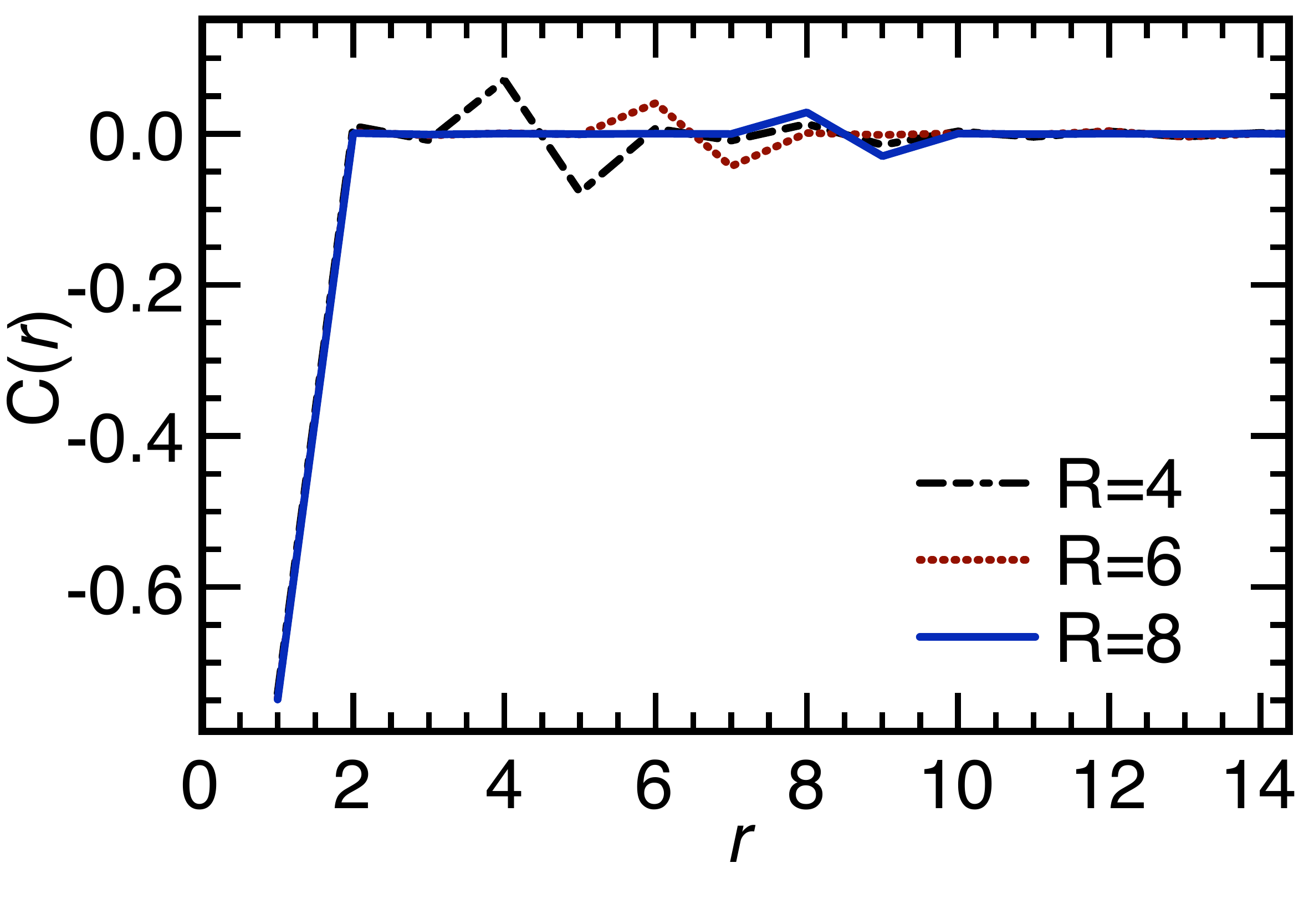}
\caption {The spin-spin correlation, $C(r)$, vs $r$.}
 \label{fig:ssc}
 \end{figure}
\begin{figure}
\centering
\includegraphics[width=8cm]{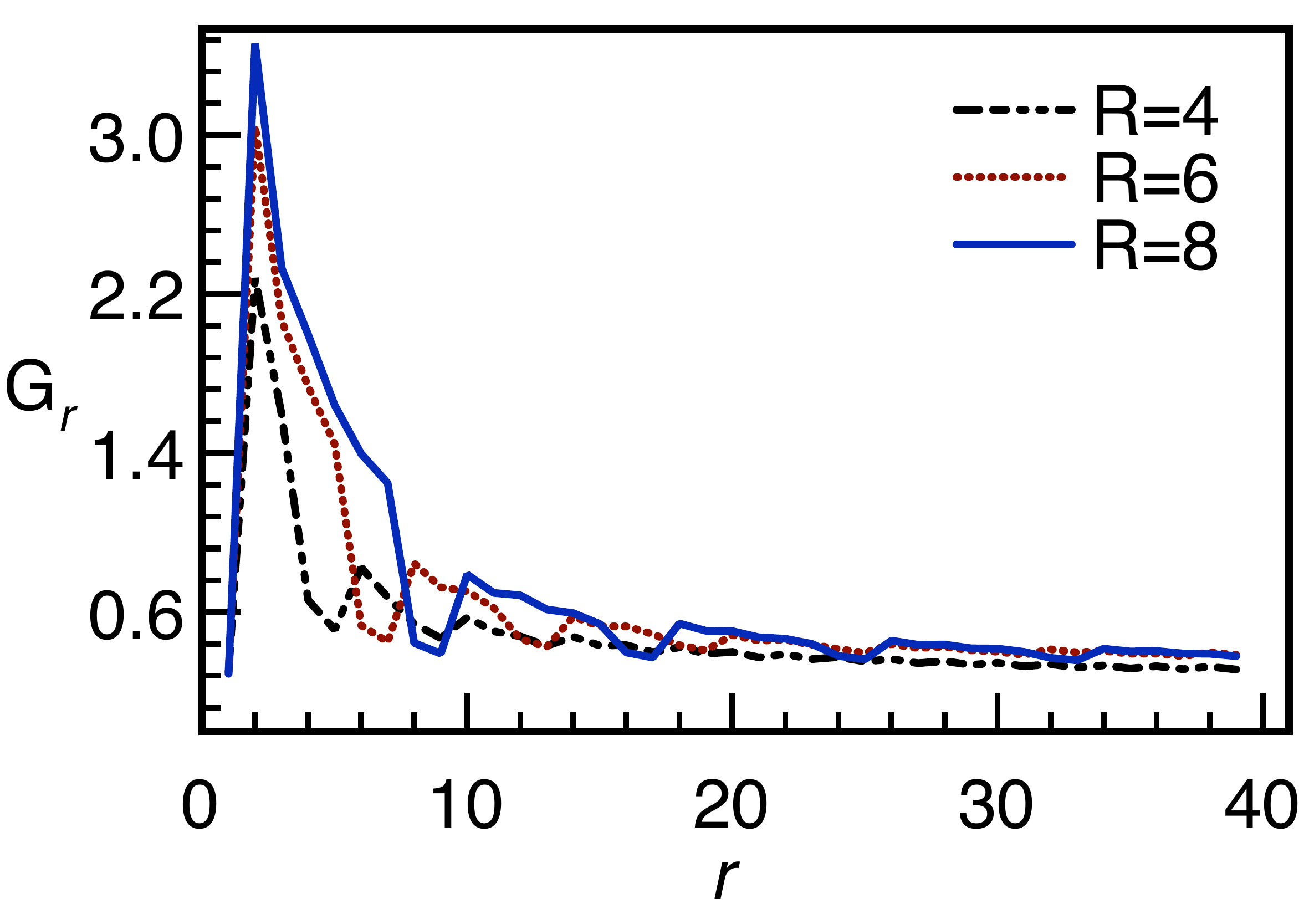}
\caption {$G_{r}=\frac{1}{r}\log[(-1)^rC(r)]$ vs. $r$  for $L=40$. }
\label{fig:Gr}
\end{figure}

We have also computed spin-spin correlation, $C(r)=\langle \S_1\cdot\S_{1+r}\rangle$, which is an important characterizer of a spin system. For better accuracy of $C(r)$, we have used finite size DMRG. The calculations have been performed on spin chains up to 100 sites long. The data for $C(r)$ is presented in Fig.~\ref{fig:ssc}. 
 
 \begin{figure}
\centering
\includegraphics[width=8cm]{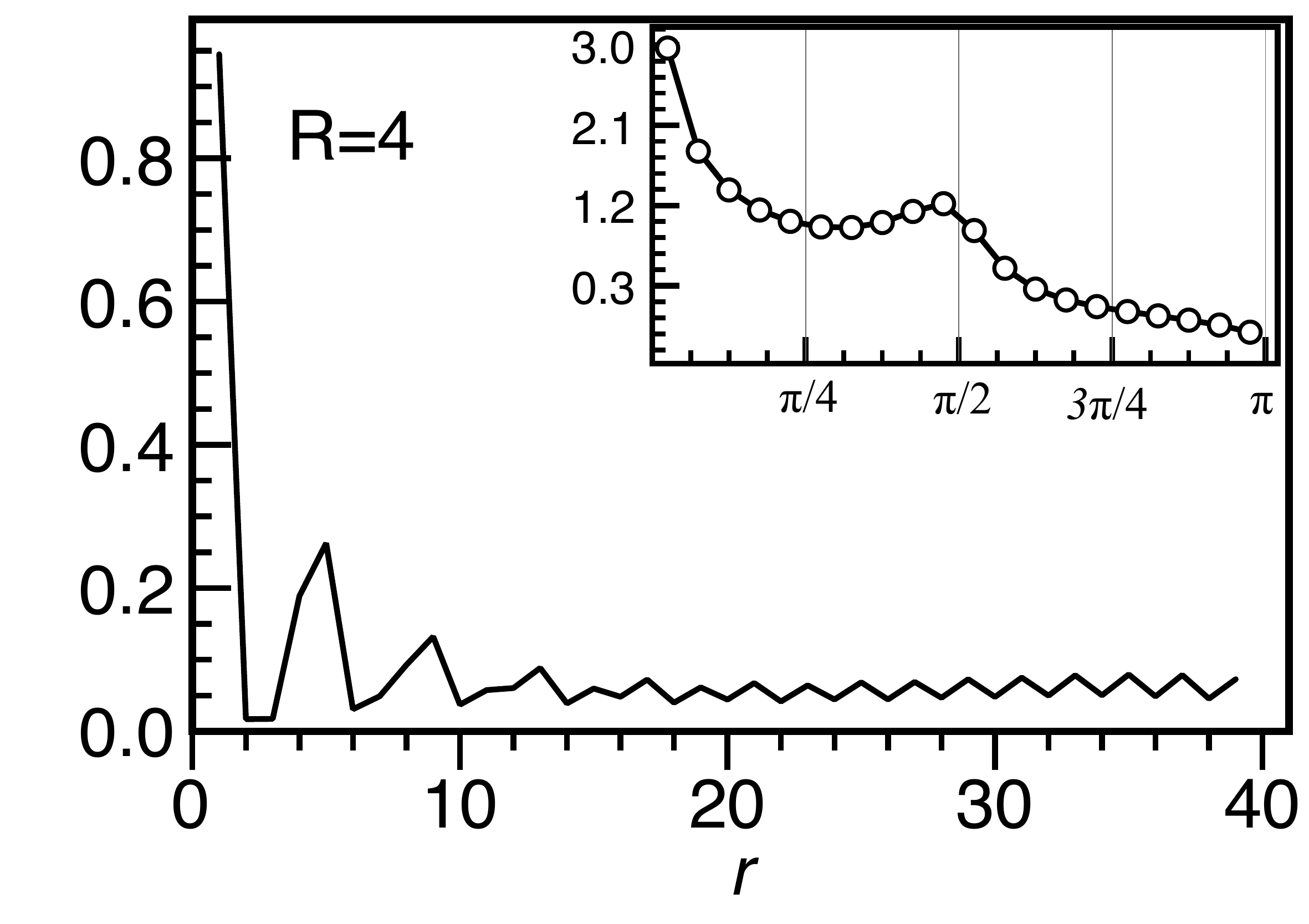}\\
\includegraphics[width=8cm]{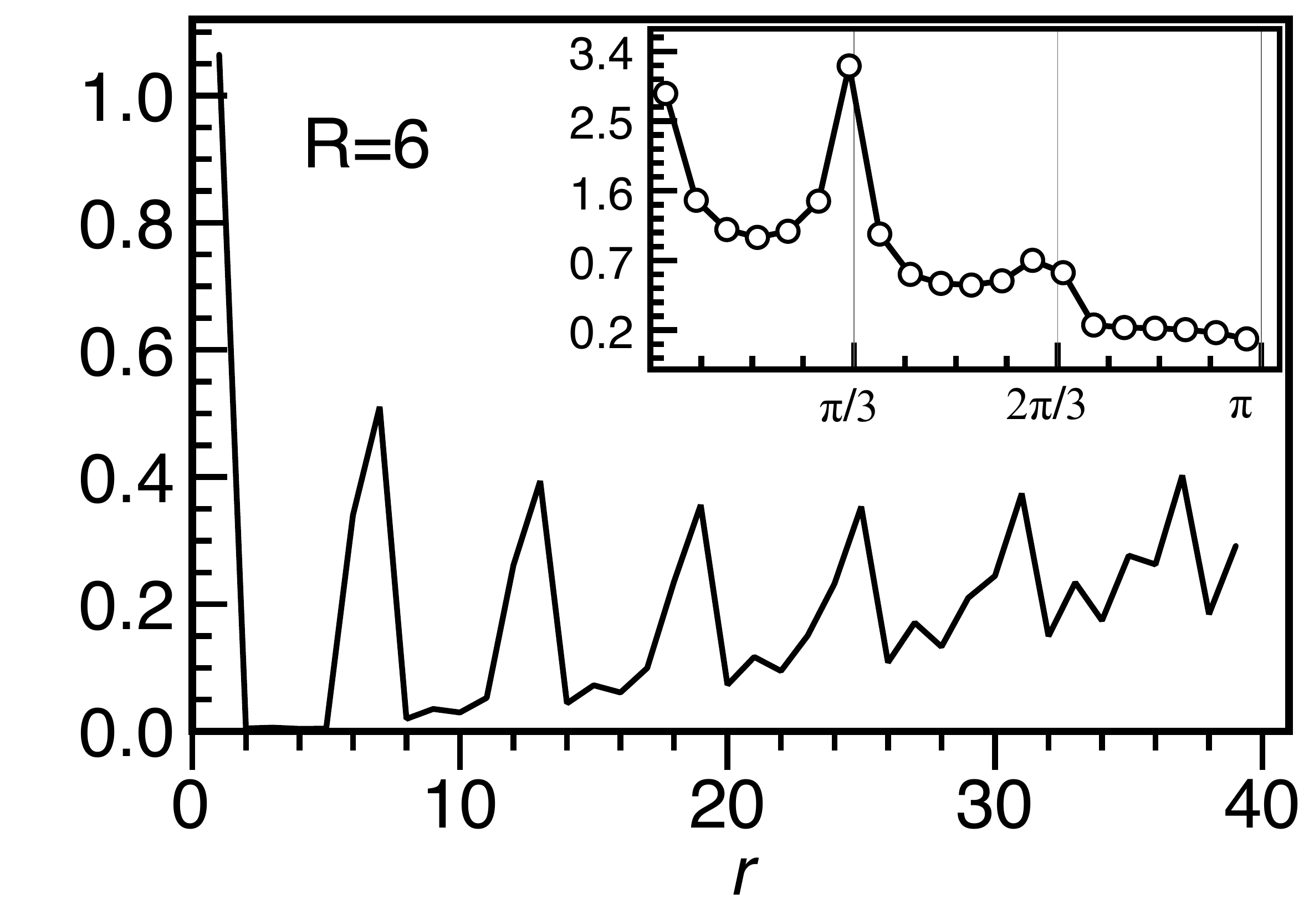}\\
\includegraphics[width=8cm]{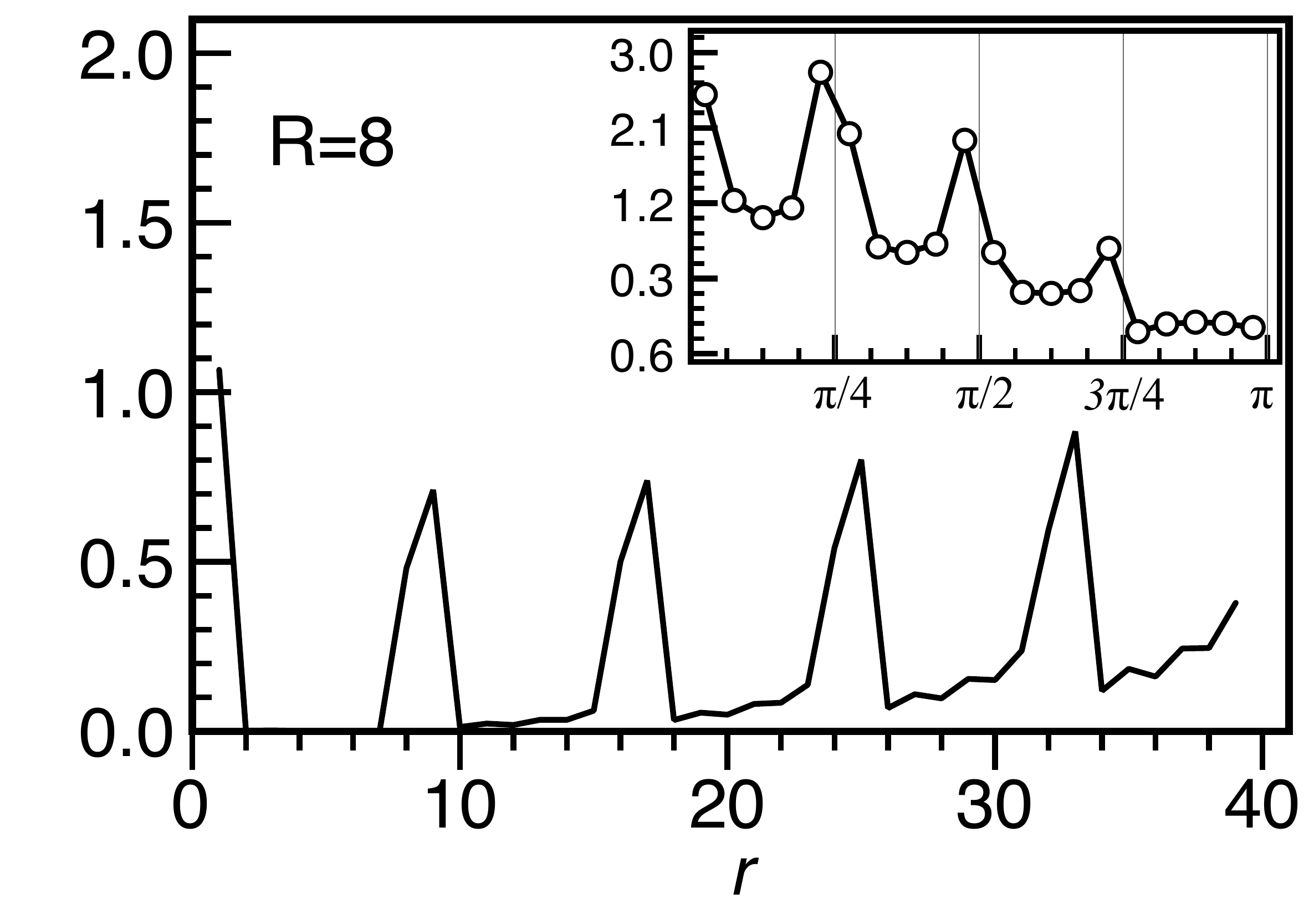}
\caption{$ (-1)^r C(r)e^{r/\xi}$ and its Fourier transform (inset).}
\label{fig:R468_corr-riv}
\end{figure}

The notable features of the spin-correlation data for even-$R\neq 2$ are the following. Unlike for $H_2$, the $C(r)$ decays very rapidly to zero upon increasing $r$. It prompts us to {\em empirically} introduce a correlation length $\xi$ such that $C(r)\sim e^{-r/\xi}$, which is consistent with the gapped behavior of the models. The sign of $C(r)$ follows $(-1)^r$, as generically expected for an antiferromagnet. We have estimated $\xi$ by computing $G_r=\frac{-1}{r}\log{[(-1)^rC(r)]}$ such that $G_\infty = 1/\xi$. Fig.~\ref{fig:Gr} clearly shows $G_r$ converging to a finite value for large $r$. The estimated numerical values of $\xi$ for $R=4$, 6 and 8 are 4.0988, 2.82876 and 2.82876, respectively. One can also estimate $\xi$ from the spin-gap as $\xi \approx 1/\Delta$.~\cite{itoi2001} Through this approximation, we find $\xi \approx 5.84604$ for $R=4$, $\xi \approx 2.83135$ for $R=6$  and $\xi \approx 2.21435$ for $R=8$. Both the estimates give comparable values of $\xi$. The accuracy for larger $R$ is lesser however. 
\subsubsection{Correlation-revival} 
\begin{figure}
\centering
\includegraphics[width=8cm]{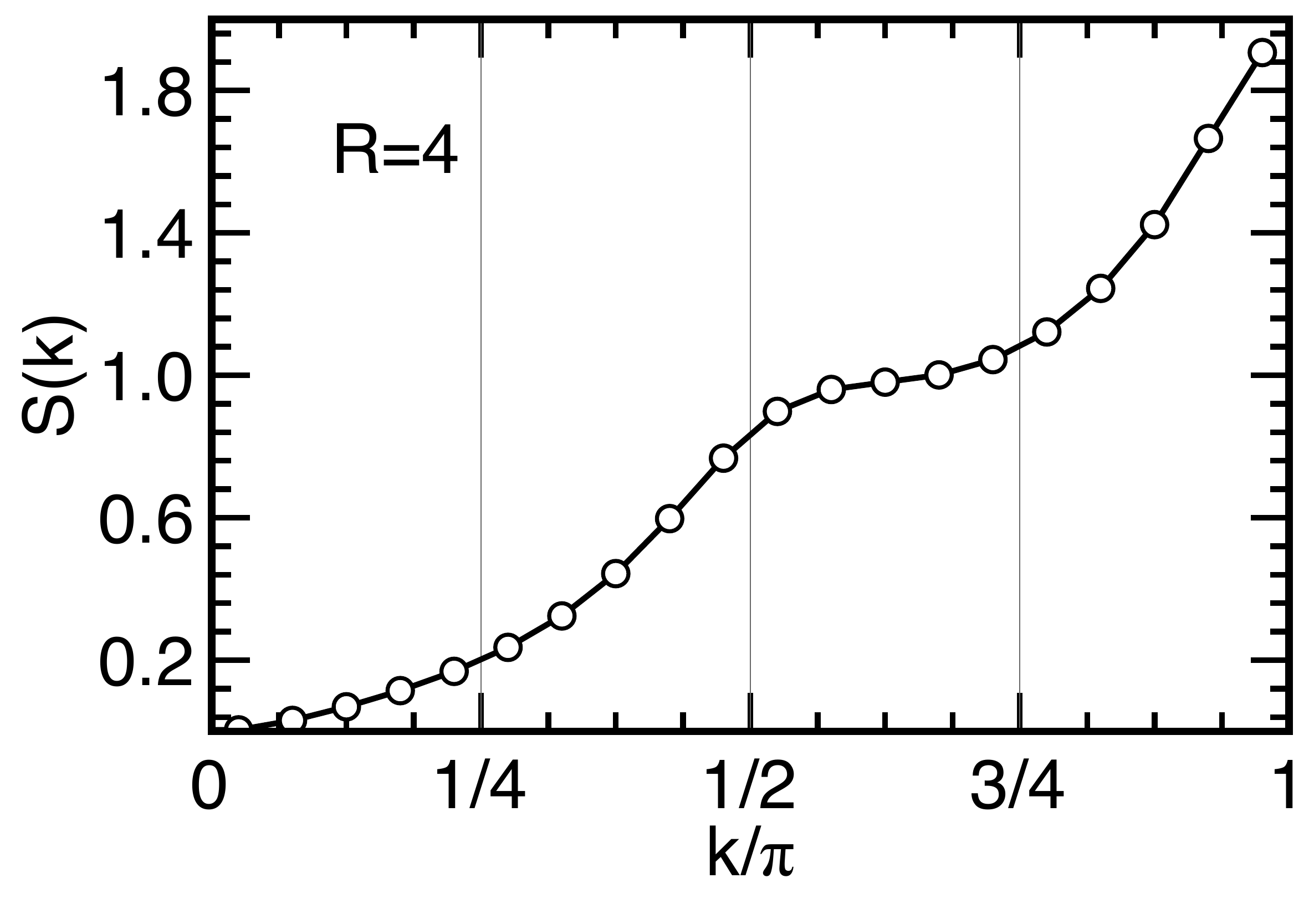}\\
\includegraphics[width=8cm]{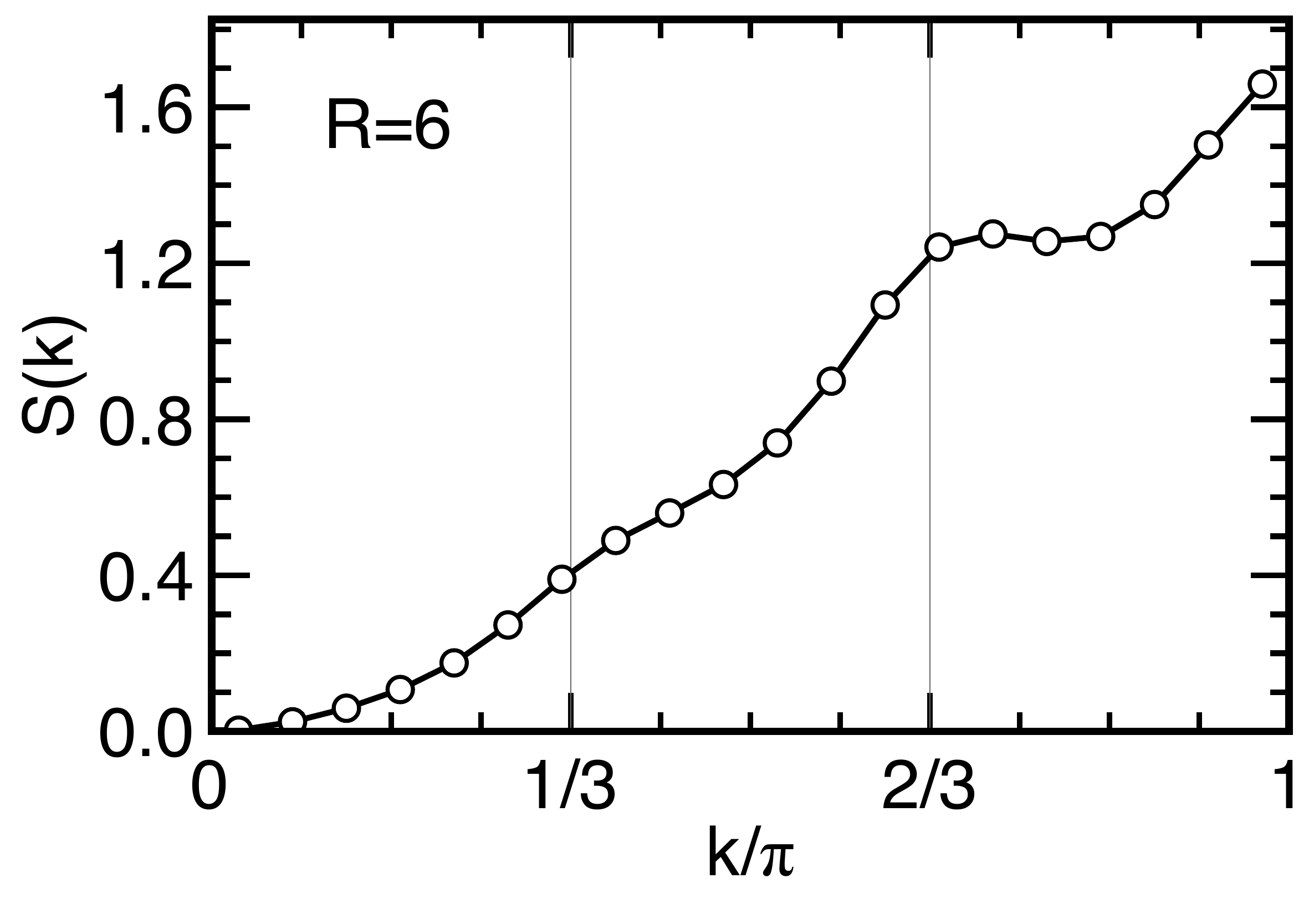}
\includegraphics[width=8cm]{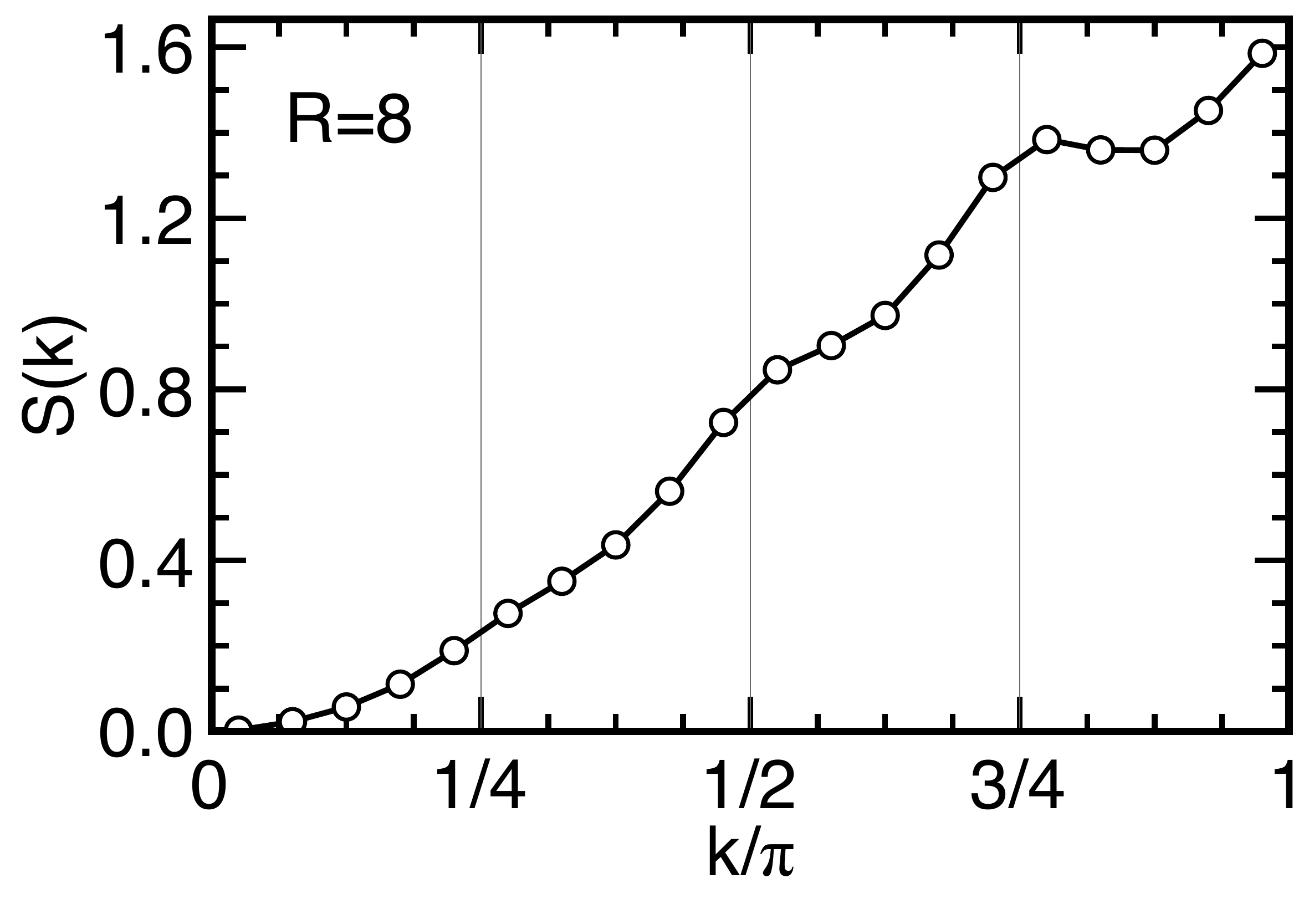}
\caption{The structure factor $S(k)$ vs $k$.}
 \label{fig:R468_Sk}
 \end{figure}
 
A careful look at $C(r)$ shows weak but repeated revivals of $C(r)$, around $r\sim R$, $2R$ and so on, of an otherwise rapidly decaying correlation function. This property of the spin correlations has been argued to be a signature of the corresponding Ising problem which has an order with enlarged unit of size $R$ (a generalized sort of N\'eel state for an even-$R$ Ising model).~\cite{bkumar-thesis} Here, we analyze this feature in some detail.
 
\begin{figure}
\centering
\includegraphics[width=8cm]{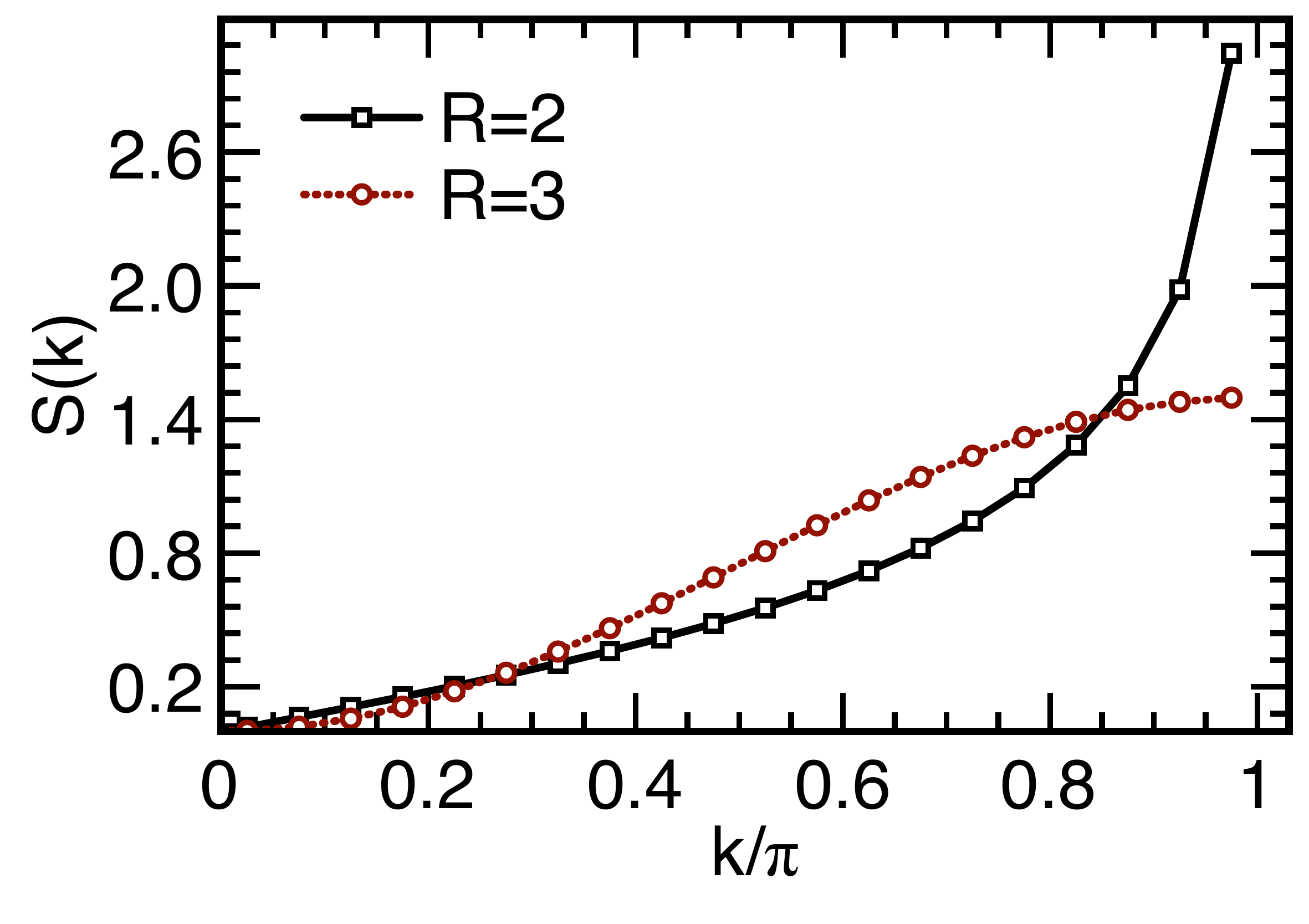}
\caption{The structure factors of the $H_2$ and $H_3$ models.}
\label{fig:R23_Sk}
\end{figure}

To clearly see the correlation-revival features, we undo the rapid decay of $C(r)$ by multiplying it with this empirical factor of $e^{-r/\xi}$, using the estimated values of $\xi$. The plots of $e^{-r/\xi}C(r)$ vs. $r$, in Fig.~\ref{fig:R468_corr-riv}, nicely show the repetitive feature. Since $R$ is roughly the period of correlation-revival, we expect $k\sim \frac{2\pi n}{R}$ (integer $n\le R/2$) to show up as special $k$-points in the corresponding structure factor, defined below.~\cite{Bursill1995}
\begin{equation}
S(k)=\frac{3}{4}+\frac{1}{L}\sum_{l=1}^{L-1}\sum_{r=1}^{L-l} C_l(r)\cos{kr}
\label{eq:Sk}
\end{equation}
Here, $C_l(r)=\langle\S_l\cdot\S_{l+r}\rangle$ is the correlation between the spins $\S_l$ and $\S_{l+r}$. The computed structure factors are shown in Fig.~\ref{fig:R468_Sk}. For any $R$, the $S(k)$ always peaks at $k=\pi$, due to the factor $(-1)^r$ in $C(r)$. In contrast to the smooth $S(k)$ for $H_2$ and $H_3$ (see Fig.~\ref{fig:R23_Sk}), there exist identifiable shoulder-like features for $H_4$, $H_6$ and $H_8$ at $k$-points other than $\pi$.
 
\begin{figure}
\centering
\includegraphics[width=8cm]{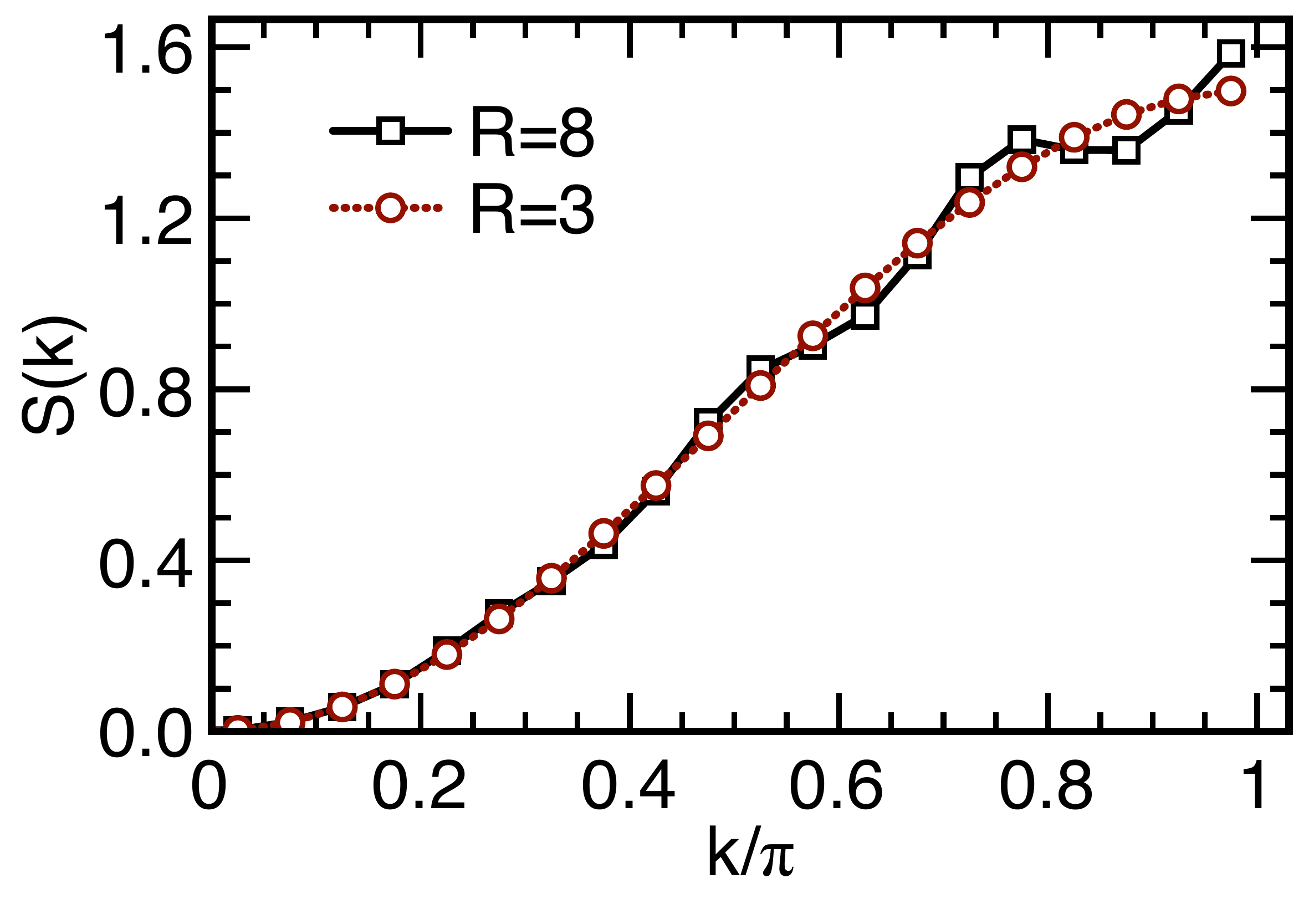}
\caption {Comparing $S(k)$ of $H_8$ with that of $H_3$ for $L=40$. The two are almost identical.}
\label{fig:Sk_H8vsH3}
\end{figure}

 As shown in Fig.~\ref{fig:R468_Sk}, the $S(k)$ for $R=4$ has a clear shoulder like feature at  $k\approx 0.525\pi$, which is not too far off from the anticipated value of $\pi/2$. In an ideally ordered situation, it would be peaked like a delta function. But because the spin correlation decays rather rapidly, this shoulder like feature is all that we can see. Similar features for $R=6$ occur around $k=\pi/3$ and $2\pi/3$. For $R=8$, they occur around $k=\pi/4$, $\pi/2$ and $3\pi/4$. So indeed, for a given even-$R$, there are $\frac{R}{2}-1$ non-trivial $k$ points as anticipated from an Ising type argument. A better view of these non-trivial $k$-points can be gained by Fourier transforming $(-1)^r e^{-r/\xi} C(r)$ as shown in the insets of Fig.~\ref{fig:R468_corr-riv}. It should also be noted that already for $R=8$, the structure factor is almost like that of $H_3$, as show in Fig.~\ref{fig:Sk_H8vsH3}.

\subsection{Entanglement}
We have also calculated some information-theoretic quantities such as the concurrence and the entropy (of entanglement) in the ground states of the even-$R$ models. These are good measures of quantum correlation. 

Since the spins in the linear-exchange models are interacting via $SU(2)$ symmetric antiferromagnetic exchange, the ground state in general is a singlet (total $S=0$ state) whether known exactly or not. For a given singlet state, the reduced density matrix of a pair of spins, obtained by tracing over the other spins, is also rotationally invariant, given by the following one parameter Werner state.~\cite{Werner1989}
\begin{eqnarray}
 \rho_{2}(i,j)&=&p[i,j\rangle \langle i,j]  +\frac {1- p} {4} I
\end{eqnarray}
Here, $[i,j\rangle$ is the singlet state formed by the spins $\S_i$ and $\S_j$, and $p=-(4/3) \big\langle\S_i.\S_j\rangle$. For $p=1$, the Werner state is a pure singlet state. In general, however, it represents a mixed state. Therefore, the entanglement between two spins is given by the concurrence, $C_{\rho^{ }_2}=\max(0,\frac {3}{2}p-\frac{1}{2})$.~\cite{Wootters1998}  

\begin{figure}
\centering
\includegraphics[width=8cm]{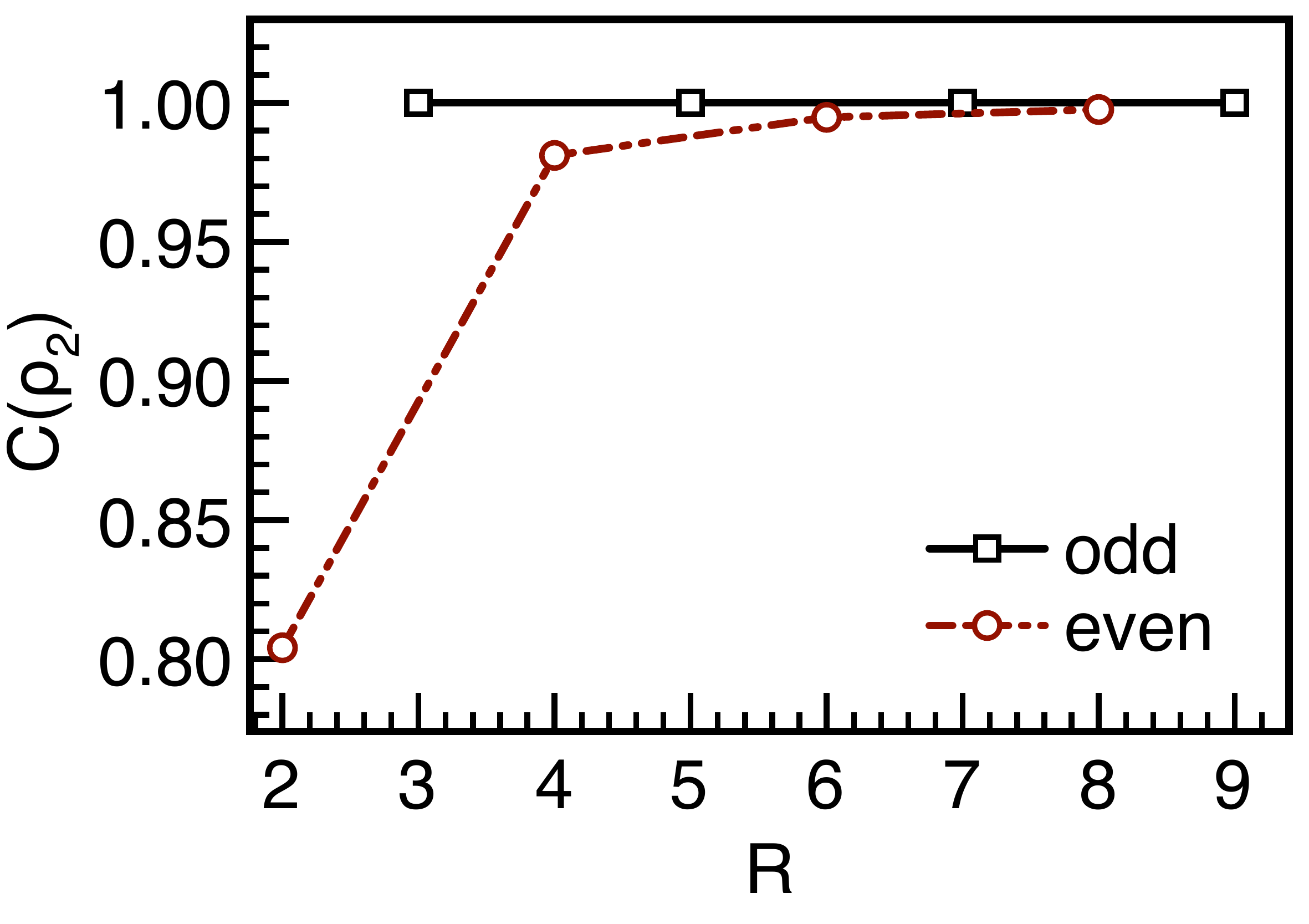}
\caption {The nearest-neighbor concurrence vs.  $R$.  }
\label{fig:R_conc}
\end{figure}

For the even-$R$ linear exchange models, we have computed $C_{\rho^{ }_2}$ for the nearest-neighbor spins for $L=40$. It is shown in Fig.~\ref{fig:R_conc}, whose strikingly similarity to the ground state energy plot (in Fig.~\ref{fig:Eg}) is immediately noticable. It is not surprising however, because the nearest-neighbor pair state tend to approach the perfect singlet as even-$R$ increases. This we know from the dimer-dimer correlations. The $C_{\rho^{ }_2}$ simply follows this approach towards the singlet state, just as the ground state energy does.

\begin{figure}
\centering
\includegraphics[width=8cm]{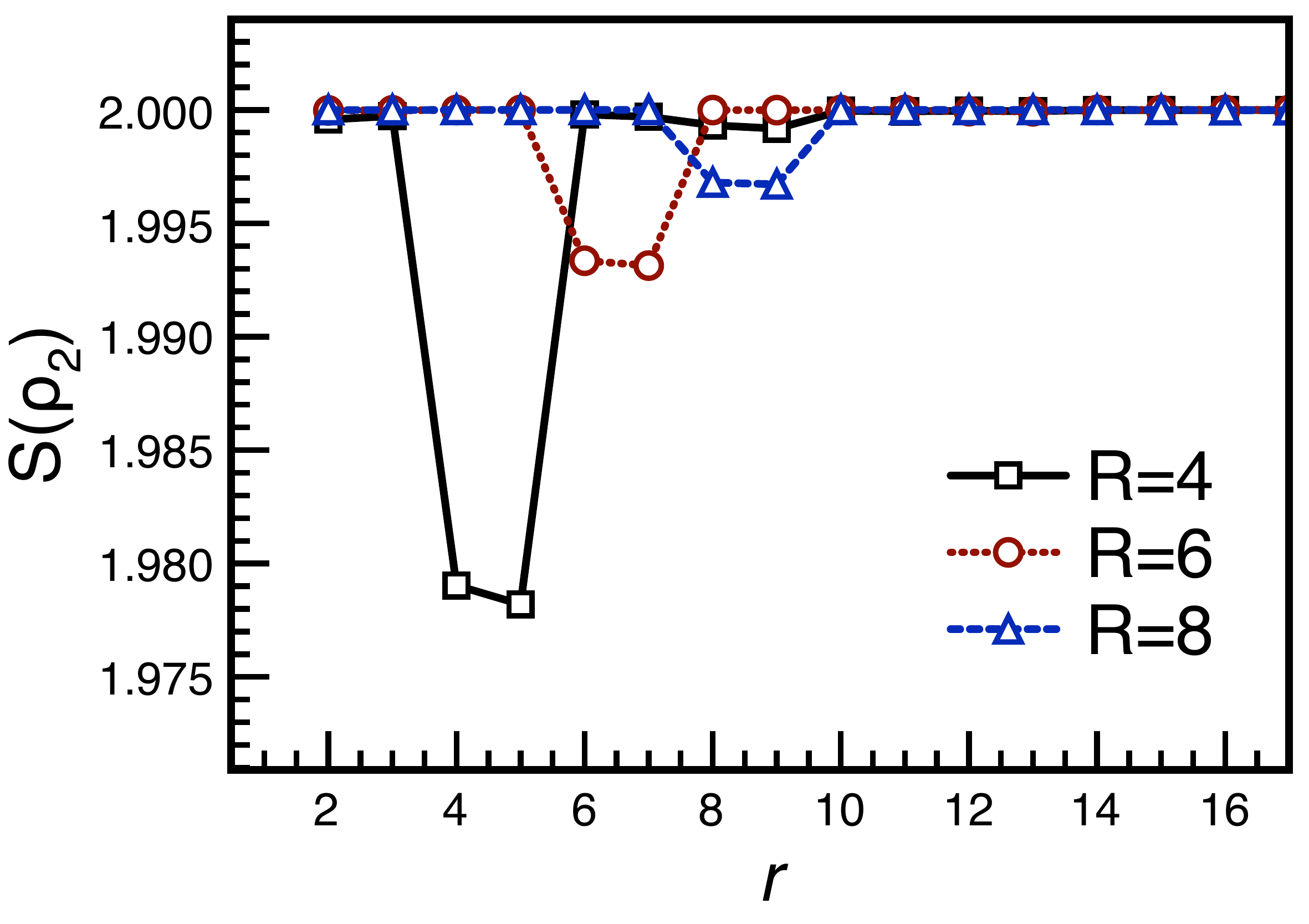}
\caption { Entanglement entropy  plotted against the separation $r$ between the spins. The values  of $S(\rho_2)$ at $r=1$, which are very close to zero, are not shown here in order to zoom into the features at $r\sim R$ that appear as small dips.}
\label{fig:entropy}
\end{figure}

The entropy is another measure of correlation within a state. For a spin-1/2 pair, if the two spins are uncorrelated, the entropy would be maximum, that is $2$ (taking $\log$ base 2). In the Werner state, it is given by the relation.~\cite{Chhajlany2007}
 \begin{equation}
S(\rho_2) = 2-\frac{1+3p}{4}\log_{2}(1+3 p) - 3\frac{1- p}{4}\log_{2}(1-p)
\end{equation}
Clearly, for a perfect singlet between a pair of spin-1/2s, that is $p=1$, the entropy $S(\rho_2)$ is zero as it ought be. Figure~\ref{fig:entropy} shows the $S(\rho_2)$ computed for different separations along the chain. Expectedly, it is close to zero for the nearest-neighbor spins. However, it quickly approaches the value of 2 for larger separations, consistent with the rapidly decaying spin-spin correlations. Interestingly, while remaining mostly equal to 2, the entropy also shows signatures of slight recovery of correlations at a distance of $R$, in accordance with the correlation-revival seen in the spin-spin correlation function. 
\subsection{Evolution of energy gap as a function of $R$}
\begin{figure}
\centering
\includegraphics[width=8cm]{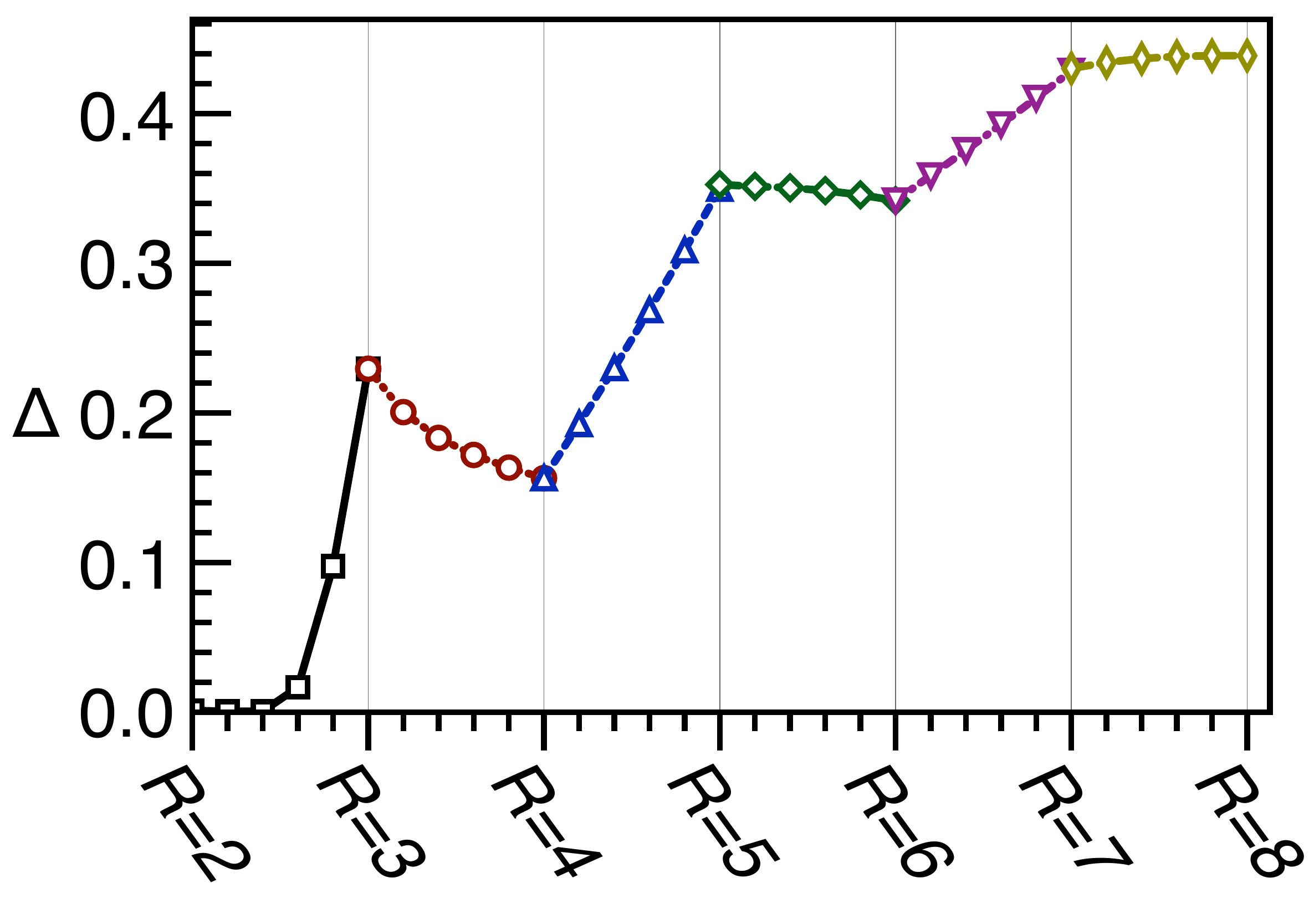}
\caption{Evolution of the spin-gap between successive linear-exchange models, as function of $\alpha$ defined in Eq.~(\ref{eq:Halpha}).}
\label{fig:gap_alpha}
\end{figure}
Consider a suitably modified version of the linear-exchange models, 
\begin{equation}
H_{R,\alpha} = H_{R} + \alpha V_{R},
\label{eq:Halpha}
\end{equation}
where $V_{R} = $ $\sum _{l} \S_{l}.\left[\S_{l+1}+\S_{l+2 }+...+\S_{l+R}\right]$, and the parameter $\alpha$ varies from 0 to 1. It has been noted~\cite{bkumar2002, bkumar-thesis} that the successive $R$ models can be related through $H_{R,\alpha}$ as: $H_{R+1}=H_{R,1}=H_R+V_R$. The $\alpha$ can thus be viewed as a parameter which continuously changes $H_R$ to $H_{R+1}$. It offers an interesting handle to see how even-$R$ evolves to odd-$R$ and vice versa. Here, we present the evolution of spin-gap as a function of $\alpha$ (see Fig.~\ref{fig:gap_alpha}). We find that the gap, $\Delta$, remains non-zero (except for $H_2$), and smoothly increases or decreases for $\alpha$ varying between 0 and 1. The spin-gap for an odd-$R$ appears as a local maxima between two even-$R$ cases.

 \section{Summary}
We have performed DMRG calculations on the linear-exchange quantum spin chains given by the Hamiltonian $H_R$ of Eq.~(\ref{eq:HR}). In particular, we have investigated the models for $R=4$, 6 and 8, in order to understand the general nature of even-$R$ models in relation to the odd-$R$ cases (whose ground state properties have been known exactly). We have computed the ground state energy, the spin-gap, the dimer-dimer and spin-spin correlations, the nearest-neighbor concurrence, and the entropy content in spin-pairs, to uncover the low-energy physics of the even-$R$ models. Based on our findings, we now confidently understand the class of linear-exchange models as summarized below.

The integrable $H_2$ stands out alone with algebraic spin-correlations in the ground state and gapless excitations. All the rest are found to be spin-gapped with rapidly decaying spin correlations.

The ground states of the even-$R\neq2$ models exhibit nearest-neighbor singlet dimerization, akin to the odd-$R$ models. In fact, numerically the case of $R=8$ already shows the almost perfect MG dimerization. The rigorous asymptotic behavior, it appears, sets in rather quickly even for not-so-very-large values of even-$R$. Thus, all the linear exchange models (except $H_2$) exhibit spontaneous dimerization like the MG model.

The one thing about even-$R\neq 2$ models that is common with $H_2$ is the $\pi$-oscillations in the spin correlation (relics of the N\'eel type ordering in the corresponding Ising cases). While the spin-correlations decay very rapidly (for $R\neq 2$), they also show weak signs of (periodic) revival, before eventually decaying to zero. In the Fourier space, these features give rise to $\frac{R}{2}-1$ identifiable $k$-points roughly close to 2$\pi n/R$ ($n=1,2,..,\frac{R}{2}-1$), as suggested by the corresponding Ising picture.    
  \begin{acknowledgments}
Bimla acknowledges CSIR-India for financial support.
\end{acknowledgments}
\bibliography{linXspin_dmrg}
\end {document}